\documentclass[%
 reprint,
superscriptaddress,
 amsmath,amssymb,
 aps,
]{revtex4-2}

\usepackage{graphicx}% Include figure files
\usepackage{dcolumn}% Align table columns on decimal point
\usepackage{bm}% bold math
\usepackage{xcolor}

\begin{document}

\preprint{APS/123-QED}

\title{A bi-effect model of muon deflections in air showers}% Force line breaks with \\
%\thanks{A footnote to the article title}%

\author{Si-Zhe Wu}
% \altaffiliation[Also at ]{School of Astronomy and Space Science, Nanjing University, Nanjing 210023, China}%Lines break automatically or can be forced with \\
\author{Chao Zhang}%
 \email{chao.zhang@nju.edu.cn}
\author{Ruo-Yu Liu}% 
 \email{ryliu@nju.edu.cn}
\affiliation{School of Astronomy and Space Science, Nanjing University, Nanjing 210023, China} 
\affiliation{Key laboratory of Modern Astronomy and Astrophysics, Nanjing University, Ministry of Education, Nanjing 210023, China}   
\author{Xi-Shui Tian}
\author{Zhuo Li}
\affiliation{Department of Astronomy, School of Physics, Peking University, Beijing 100871, China} 
\affiliation{Kavli institute for astronomy and astrophysics, Peking University, Beĳing, China}
%This line break forced with \textbackslash\textbackslash

%\collaboration{MUSO Collaboration}%\noaffiliation

%\author{Charlie Author}
% \homepage{http://www.Second.institution.edu/~Charlie.Author}
%\affiliation{
% Second institution and/or address\\
% This line break forced% with \\
%}%
%\affiliation{
% Third institution, the second for Charlie Author
%}%
%\author{Delta Author}
%\affiliation{%
% Authors' institution and/or address\\
% This line break forced with \textbackslash\textbackslash
%}%
%\collaboration{CLEO Collaboration}%\noaffiliation

\date{\today}% It is always \today, today,
             %  but any date may be explicitly specified

\begin{abstract}
Recent progress has shown that the geomagnetic field exerts a more significant impact than expected on the behavior of charged secondary particles in inclined air showers. In this study, we for the first time combine it with atmospheric effects to construct a bi-effect model, aiming to investigate the lateral distribution of particles on the ground plane. Despite the complex physical interactions during the development of air showers, a simple formula can describe the overall deflection of $\mu^{\pm}$ and accurately fit the deflection in simulated air showers, thereby validating the hypotheses about these effects in this study. Furthermore, we have obtained the relationship between model parameters and primary particle information for different experimental sites. This new model is highly successful and is promising to provide new insights for improving detector layout design and air shower reconstruction.
% which could be reflected by the core deflection in the particle distribution, and introduce challenges in the reconstruction of inclined cosmic rays.

%\begin{description}
%\item[Usage]
%Secondary publications and information retrieval purposes.
%\item[Structure]
%You may use the \texttt{description} environment to structure your abstract;
%use the optional argument of the \verb+\item+ command to give the category of each item. 
%\end{description}
\end{abstract}

%\keywords{Suggested keywords}%Use showkeys class option if keyword
                              %display desired
\maketitle

%\tableofcontents

%\section{\label{sec:level1}Introduction}
%\paragraph
\section{Introduction}
%\emph{Introduction —} 
Cosmic rays (CRs) are high-energy particles originating from outer space. They serve as probes to study extreme environments in astrophysics and particle physics at energies beyond those achievable by human-made colliders. Many important questions are expected to be answered through the study of cosmic rays. In particular, the origin and nature of ultra-high-energy cosmic rays (UHECRs) remain mysterious \cite{Bhattacharjee:1999mup, AlvesBatista:2019tlv}. Extensive air showers — massive cascades of secondary particles — are triggered when cosmic rays collide with atomic nuclei in the atmosphere and can be observed via large detector arrays. Observatories such as the Pierre Auger Observatory, LHAASO, and TA are working to collect data, and several hypotheses are being tested to provide reasonable explanations \cite{PierreAuger:2017pzq, LHAASO:2019qtb, TelescopeArray:2012qqu, PierreAuger:2016qzj}. 

The secondary particles of air showers consist of a rich mixture of hadrons, electromagnetic particles, muons, and neutrinos. These particles propagate through Earth's atmosphere during the development of air showers. Among them, the electromagnetic components ($e^{\pm}, \gamma$)  and muons ($\mu^{\pm} $) play a crucial role in detection. Although electromagnetic particles carry most of the energy deposited in the atmosphere, they undergo more intense interactions and are thus more absorbed. Especially, in inclined air showers, electromagnetic particles are much harder to reach the ground compared to vertical air showers.
Muons are produced from the decay of charged pions. They interact weakly with atomic nuclei in the atmosphere and can persist until the final stage of air shower development. As a result, they have low energy loss and a long lifetime, enabling them to penetrate the atmosphere and reach the ground, underground, or deep water. The technology for detecting muons is quite mature; their distribution and energy spectrum can be accurately measured using surface detectors, making muon detection an important method for cosmic ray research \cite{PierreAuger:2021qsd}.

On one hand, muon detection can serve as a probe to study hadronic interactions in air showers, distinguish between heavy and light primary nuclei, and reconstruct the energy of cosmic rays. On the other hand, many muon-related challenges remain unsolved, such as the model-data discrepancy in muon number measurements — the so-called ``muon puzzle'' \cite{Pierog:2017awp, Pierog:2025ixr}; the precise determination of mass composition using muon data \cite{Knurenko:2020qvz, Mayotte:2025cqr}; and the separation of muon signals from electromagnetic particles in detectors \cite{PierreAuger:2021nsq}. Additionally, muon flux is significantly influenced by environmental factors such as detector altitude and solar modulation \cite{Bae2021frw}.
In previous studies, atmospheric effects have been considered to investigate particle distributions. A universality model has been established to accurately describe the longitudinal, lateral, and temporal distributions of particles, which could be applied to study mass composition \cite{Stadelmaier:2024pae}. This model has been applied to real experimental data \cite{Stadelmaier:2025pqd}. However, most previous studies are limited to zenith angles not exceeding 60$^{\circ}$.

Recent progress in radio emission mechanisms suggests a new radiation effect induced by the movement of charged particles in a relatively high geomagnetic field strength within inclined air showers \cite{Chiche:2024yos}, indicating that the charged secondary particles should also be more influenced by the magnetic field than expected. However, this effect has not been investigated in air shower detection, except for the balloon-based EUSO-SPB2 Mission \cite{Fuehne:2023kap}.

This paper aims to study the behaviors of muons reaching ground level by combining atmospheric and geomagnetic effects. Additionally, to make this model generic, both vertical and inclined air showers have been taken into consideration. To simplify the analysis, the averaged coordinates of muons on the ground plane are used to represent the muons' behaviors.

%The extensive air showers induced by cosmic rays produce many types of particles, on ground level, it contains mainly photons, e$^{\pm}$ and $\mu^{\pm}$, which carry normally more than 80\% of the energy of the shower. Discovery of high-energy cosmic rays, which are based on the Cherenkov effect, the radio emission in air showers including the transverse current and Askaryan effect of secondary particles in the shower, has inspired the invention of particle detectors, and brought breakthroughs to our understanding of elementary particle. The success of radio detection of cosmic rays is one of the profit we have obtained in this kind of work. (to continue)
%\paragraph

\section{The bi-effect model}
\begin{figure}[htbp]
    \includegraphics[width=0.8\linewidth]{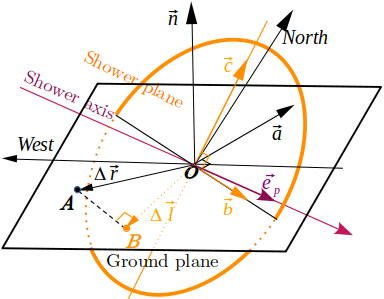}
\caption{\label{fig:epsart} Coordinate system. The point where the shower axis intersects the observation plane is defined as the origin $O$. $\vec{e_p}$ represents the unit vector in the cosmic ray direction. The unit vectors ($\vec{b}$, $\vec{a}$, $\vec{n}$) and ($\vec{c}$, $\vec{b}$, $\vec{e_p}$) form right-handed coordinate systems for the ground plane and shower plane, respectively.}
\label{coordinates}
\end{figure}

\emph{Coordinate system —}
%\label{build model}
%Define the coordinate system as FIG.\ref{atmosphere_absorption}. 
This study employs the  Cartesian coordinate system  defined in CORSIKA\cite{Heck1998vt}, where the origin is set to the intersection point of the air shower axis with the ground plane. The $x$-axis points to the magnetic north, the $y$-axis to the west, and the $z$-axis upward. A schematic diagram is shown in  FIG. \ref{coordinates}. By counting the deflection of each particle on the ground, the overall deflections of particle number and energy are defined respectively as follows:  
%The deflection $\Delta{\vec{r}}$ of the particles on the ground plane can be calculated directly. 
% $\vec{r_N}$ and $\vec{r_E}$:
\begin{equation}
\begin{matrix}
 \Delta{\vec{r}_N}= \frac{\sum_i\Delta{\vec{r}_i}}{n} &,&  \Delta{\vec{r}_E}= \frac{\sum_i E_i\Delta{\vec{r}_i}}{\sum_i E_i}
\end{matrix}
\end{equation}
where $\Delta{\vec{r}} $  denotes the distance to the origin, $E_i$ denotes the energy of each particle. 
%As the power-law energy distribution of secondary particles, $\Delta{\vec{r_N}}$ mainly reflects the properties of low-energy secondary particles, while $\Delta{\vec{r_E}}$ mainly reflects high-energy particles.

\emph{Model hypotheses — } The energy loss of muons is proportional to the amount of matter they encounter in the atmosphere. The energy loss rate of muons due to ionization is relatively constant at 2 MeV/$g \cdot cm^{-2}$ \cite{Barazandeh:2016lhh}. Meanwhile, as charged particles, muons have their trajectories deflected by the Lorentz force in the geomagnetic field. In particular, muons in inclined air showers propagate for a longer time than those in vertical air showers. When both the deflection caused by the magnetic field and atmospheric absorption are taken into account, two core hypotheses can be used to construct the physical model:
%($E=E_0 \cdot e^{-x/X_0}$).
 %These hypotheses will be validated in Section \ref{Validation of hypothesis}. 

1. Independence Hypothesis: We assume that the magnetic deflection  ($\Delta \vec{r}_{mag}$) caused by the Lorentz force and the atmospheric absorption-induced displacement ($\Delta{\vec{r}_{atm}} $) are independent. This relationship is expressed as: $\Delta{\vec{r}}= \Delta{\vec{r}_{mag}}+\Delta{\vec{r}_{atm}} $.

2. Magnetic Linearity Hypothesis: Charged particles undergo Larmor motion in the geomagnetic field. When the change in the direction of particle motion is minimal, we approximate the direction of the Lorentz force as constant; under this condition, Larmor motion is approximated as projectile motion. The deflection of muons in the shower plane $\Delta{\vec{l}}$  is proportional to the Lorentz force, which is expressed as: $ \Delta{\vec{l}_{mag}} = c_{mag}\ \vec{e}_p\times \vec{B}$ where $\vec{B}$ denotes the magnetic field at the observation site, and the parameter $c_{mag}$  is a constant independent of both the azimuth angle of the primary particle and the magnetic field.

\emph{Deflection model —}
We simply project the $\Delta{\vec{l}_{mag}}$ onto the ground plane to obtain:  $\Delta{\vec{r}_{mag}}$:
\begin{equation}
\begin{split}
\Delta{\vec{r}_{mag}} &= (\vec{b} \cdot \Delta{\vec{l}_{mag}})\ \vec{b}\ -\ \frac{\vec{c}\cdot\Delta{\vec{l}_{mag}}}{\vec{e}_p\cdot\vec{n}}\ \vec{a} \\ 
 &=c_{mag}\left [ \vec{b} \cdot (\vec{e}_p\times \vec{B})\ \vec{b}\ -\ \frac{\vec{c}\cdot(\vec{e}_p\times \vec{B})}{\vec{e}_p\cdot\vec{n}}\ \vec{a} \right ] 
\label{twoeffects} 
\end{split}
\end{equation}

where $c_{mag}$ is the magnetic reflection parameter, $\Delta{\vec{l}}$ and $\Delta{\vec{r}}$ denotes the deflection in the shower plane and on the ground plane, respectively.  

\begin{figure}[htbp]
    \includegraphics[width=1\linewidth]{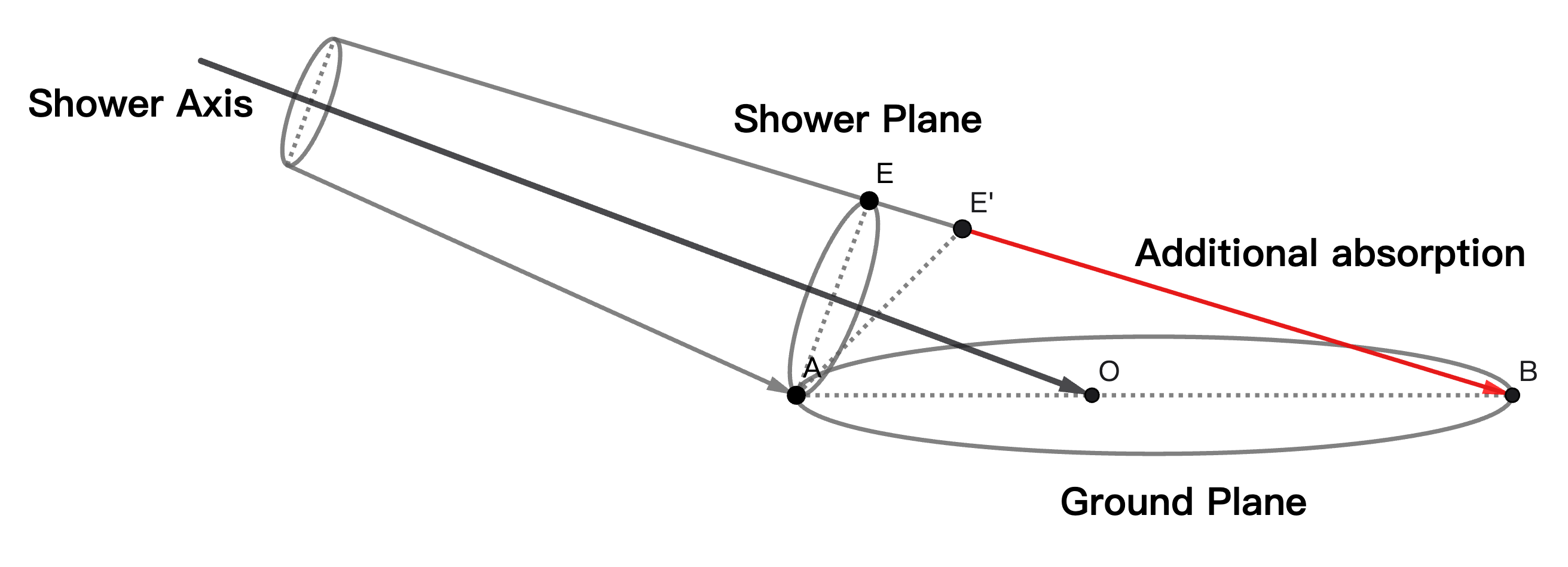}
\caption{\label{fig:epsart} Atmospheric absorption diagram. When primary cosmic ray is slanted into the atmosphere, different places on the ground or shower plane have different depths. Particles at point A and E$'$ have same atmosphere depth.}
\label{atmosphere_absorption}
\end{figure}

As illustrated in FIG.\ref{atmosphere_absorption}, the absorption of  particles above the shower axis  is more severe than that below the shower axis, resulting in fewer particles and lower energies on the side of point B than point A. This creates an additional atmospheric deflection $\Delta \vec{r}_{atm}$ in the direction opposite to $\vec{a}$, with an absolute value $|\Delta \vec{r}_{atm}|$  independent of the azimuth angle. This deflection could be expressed as $ \Delta \vec{r}_{atm}=c_{atm}\ \vec{a} $, and $c_{atm}$ is a negative constant (independent of the  azimuth angle)  representing the atmospheric deflection parameter. 

%In addition, because the particles on the B side take more time to reach the ground, the effect of the decay of low-energy muons cannot be ignored either, but the deflection caused by decay has the same properties as it caused by atmosphere (with the same direction, and the magnitude is also independent of the azimuth), so actually $c_{atm}$ has already takes into account the effect of decay.

Finally, by substituting $\Delta \vec{r}_{mag}$ and $\Delta \vec{r}_{atm}$ into Eq.\ref{twoeffects},  the total deflection is given by:

\begin{multline}
%\begin{split}
    \Delta{\vec{r}} = c_{mag}\left [\ \vec{b} \cdot (\vec{e}_{p}\times \vec{B})\  \right ]\ \vec{b}\ \\ 
    +\left [\ c_{atm}\ -\ c_{mag}\ \frac{\vec{c}\cdot(\vec{e}_{p}\times \vec{B})}{\vec{e}_{p}\cdot\vec{n}}\right ]\ \vec{a}
%\end{split}
\end{multline}

In this model, given the direction of the primary particle  and the magnetic field, once the value of two deflection parameters $c_{mag}$, $c_{atm}$ are known,  the deflection vector $\Delta{\vec{r}}$ can be calculated.

\emph{Air shower simulations —} To validate the model hypotheses and establish the relationship between ($c_{mag}$, $c_{atm}$) and the physical parameters of primary particle, a library of air showers was prepared using  CORSIKA-77550 \cite{Heck1998vt}. 

To test the impact of the deflection in different experiments, the magnetic fields of LHAASO, SKA, and the Pierre Auger Observatory were selected. 
For each detector, as presented in Table \ref{tab:my_label},  air showers were generated by performing simulations with different parameter combinations, including particle type, primary energy, zenith angle, azimuth angle, and first interaction point (FIXHEI).  The Linsley's standard model of atmospheric refractive index was applied in this work \cite{MITRA2020102470}. Together with several hundreds showers for testing, a total of approximately 11,000 air showers were prepared.
%(Additionally, several sets of simulations with different energy and magnetic fields were also conducted.)  

\begin{table}
    \centering
    \begin{tabular}{l|l} %\begin{tabular}{!{\vrule width 1pt} c | c !{\vrule width 1pt}}
        \hline
        \textbf{N$^\circ$ of showers}  & 11000 \\ \hline
        \textbf{Primaries} & p [50\%], Fe [50\%] \\ \hline
        \textbf{Energies E  [$\mathrm{eV}$]}
        & 10$^{14}$, 4$\times$10$^{14}$, ...  1.024 $\times$10$^{17}$\\ 
        \hline
        \textbf{Zenith angles}  & 50$^\circ$ , 60$^\circ$ , ... , 80$^\circ$  \\ 
        \hline
        \textbf{Azimuth angles}  & 0$^\circ$ , 5$^\circ$ , ... , 355$^\circ$  [72 steps]\\ 
        \hline
        \textbf{Hadronic models}  & QGSJETII-04, URQMD1.3cr \\
        \hline 
        \textbf{Thinning $\epsilon_{\mathrm{thin}}$} & 1 $\times 10^{-5}$ \\ 
        \hline
        \textbf{Atmospheric model} & Linsley's standard model \\ \hline
        \textbf{FIXHEI} & 10, 15, 20, ..., 35km \\ \hline
        \textbf{B field, LHAASO site} & \begin{tabular}[c]{@{}l@{}}$B_{tot}$: 55.997 $\mu$T, at 4424 m \\ Inc: 60.79$^\circ$,  Dec: 0.36$^\circ$\end{tabular} \\ \hline
        \textbf{B field, SKA site} & \begin{tabular}[c]{@{}l@{}}$B_{tot}$: 55.607 $\mu$T, at 1000 m \\ Inc: -60.02$^\circ$,  Dec: 0.1$^\circ$\end{tabular} \\ \hline  
        \textbf{B field, AUGER site} & \begin{tabular}[c]{@{}l@{}} $B_{tot}$: 23.474 $\mu$T, at 1400 m \\ Inc: 37.30 $^\circ$,  Dec: 0.12$^\circ$\end{tabular} \\ \hline
    \end{tabular}
    \caption{Parameters of the air shower library generated by the CORSIKA simulations.}
    \label{tab:my_label}
\end{table}

\section{Model validation and analysis}
\begin{figure*}
\includegraphics[width=0.8\textwidth]{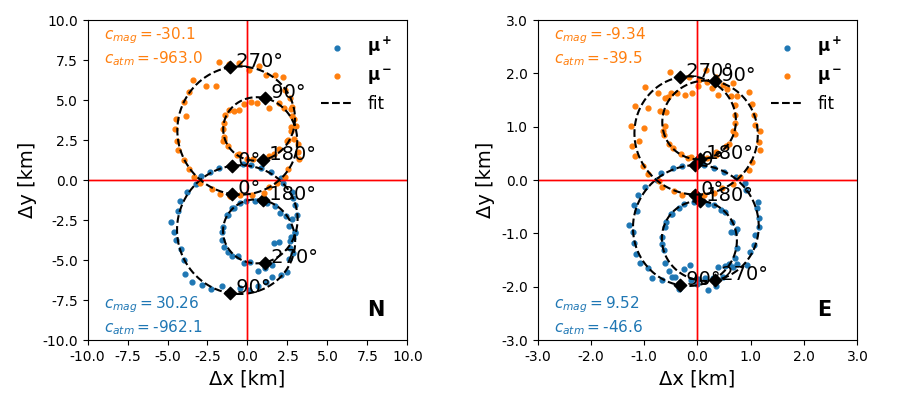}% Here is how to import EPS art
\label{deflection1}
\caption{Deflection model V.s simulations with the LHAASO parameters, and the species, zenith, altitude of first interaction points (FIXHEI), energy of primary particle is proton, 80$^\circ$, 25 km, 1e7 GeV. This dots and dashed curves represent the simulations and fitting curves, respectively. The overall deflection of particle number N (left) and energy E (right) show similar behaviors, but with different deflection parameters ($c_{mag}$, $c_{atm}$). The numbers on the curves represent some azimuth angles.}
\end{figure*}

\emph{Model validation — } 
The parameters of the  LHAASO site ($B_x=35.1  \mu T$, $B_z=-35.9  \mu T$) are selected for validation. In the model hypotheses, $c_{mag}$ and $c_{atm}$ are independent of the azimuth angle. As shown in the FIG. \ref{DC}, when the magnetic deflection parameter $C_{mag}$ is set to 0, the deflection model fits the simulation results well across different azimuth angles and effectively reproduces the double-circle structure of the magnetic deflection — validating both the hypotheses and the azimuth angle independence of the parameters.  Additionally, all the primary particle parameters are fixed, while only the azimuthal angle is varied from $0^\circ$ to $355^\circ$ in $5^\circ$ increments, resulting in a total of 72 simulations.  As shown in FIG. \ref{deflection1}, using a single pair of $c_{mag}$, $c_{atm}$, the model effectively reproduce the double-circle structure of the magnetic deflection, further validating the hypotheses and the azimuth angle independence.

Since $c_{atm}(\mu^+)\simeq c_{atm}(\mu^-)$, and $c_{mag}(\mu^+)\simeq -c_{mag}(\mu^-)$, the deflection curve of $\mu^+$ and $\mu^-$ are essentially symmetric along the north-south direction, which further confirms the hypotheses.

%\begin{figure}[htbp]
%    \centering
%    \begin{subfigure}
%        \includegraphics[width=0.4\textwidth]{DC/20241266/fit of antimuon_N.png}
%        \caption{$\Delta\vec{r_N},\  \mu^+$\label{1266_+_N}}
%    \end{subfigure}
%    \begin{subfigure}
%        \includegraphics[width=0.4\textwidth]{DC/20241266/fit of antimuon_E.png}
%        \caption{$\Delta\vec{r_E},\  \mu^+$\label{1266_+_E}}
%    \end{subfigure}
%    \caption{The deflection vector and fitting curve of $\mu^+$ and $\mu^-$. Blue %dot in fitting curve represent the Azimuth angle [$^\circ$]. The species, zenith, fixhei, energy of primary particle is proton, 80$^\circ$, 25 km, 1e7 GeV. The detector is LHAASO (4424 m, $B_x=35.1\mu T$, $B_z=-35.9\mu T$). }
%    \label{DC_1266}
%\end{figure}

%\begin{figure}[htbp]
%    \includegraphics[width=1.2\linewidth]{DC/20241266/merge.png}
%\caption{\label{fig:epsart} Atmospheric absorption diagram. When primary %cosmic ray is slanted into the atmosphere, different places on the ground or %shower plane have different depths. Particles at point A and E$'$ have same %atmosphere depth.}
%\label{atmosphere_absorption}
%\end{figure}

%We further validated the independence hypothesis through several sets of simulations in which the geomagnetic field was removed as shown in FIG.\ref{nomag}. The line of $c_{atm}(With \ B)=c_{atm}(No \ B)$ is essentially within the $3\sigma$ range, so we can basically conclude that $c_{atm}$ is independent of the magnetic field.

\emph{Fitting formula for deflection parameters —}
For each set of simulation parameters, there are corresponding values of  $c_{mag}$, $c_{atm}$ — this enable the investigation of the relationship between them.
By fitting the particle deflection data using these parameters, $c_{mag}$ and $c_{atm}$ are expected to be expressed as functions of the parameters of the primary particles.

%Next, we will attempt to establish the relationship between ($c_{mag}$, $c_{atm}$) and the physics parameters of primary particle. Once the expression is obtained, we can link the parameters of primary particle with the deflection, allowing us to infer the incident angle, particle type, and other information of the primary particle based on the deflection pattern of secondary particles.

%\emph{Atmosphere model and $\Delta$ depth}
Due to the inhomogeneity of atmospheric density,  atmospheric absorption is proportional to the atmospheric depth between the first interaction point (FIXHEI) and the observation level (OBSLEV). This atmospheric depth is defined as 
$ \tau=  \frac{\int_{\mathrm{OBSLEV}}^{\mathrm{FIXHEI}} \rho(h)dh}{\cos\theta} $
where $\rho(h)$ denotes the atmospheric density at altitude h, derived from the Linsley standard model and calculated using GDASTOOL \cite{MITRA2020102470}. 

To align with physical reality, the fitting function for $c_{mag}$ must satisfy the following boundary conditions:
\begin{equation}
\left\{\begin{matrix}
\theta \to 0^\circ, & |\Delta\vec{r}_{mag}|> 0 & \mathrm{(i)}\\
\theta \to 90^\circ, & |\Delta\vec{r}_{mag}|\to \infty & \mathrm{(ii)} \\
\mathrm{FIXHEI}\to \mathrm{OBSLEV}, & |\Delta\vec{r}_{mag}|\to 0 & \mathrm{(iii)}\\
\mathrm{FIXHEI}\to 112 km , & |\Delta\vec{r}_{mag}|> 0 & \mathrm{(iv)}
\end{matrix}\right.
\end{equation}

These conditions represent the following, respectively:\\
(i) When a primary particle enters the atmosphere vertically from above, the presence of a transverse magnetic field can still cause the deflection.\\
(ii) When a primary particle is incident parallel to the ground, the shower core is located at infinity, making it impossible for secondary particles to reach the vicinity of the origin.\\
(iii) When a primary particle undergoes a reaction only at the observation plane, no air shower is generated.\\
(iv) The maximal deflection should be limited by the thickness of the atmosphere.
%In Eq.\ref{atmosphere model}, the thickness of the atmosphere is finite. 
%Therefore, when $\theta < 90^\circ$, the distance traveled by secondary particles in the atmosphere, as well as the resulting deflection, should also be limited.
Accordingly, one formula that satisfies these conditions and provides a good fit is given by:
\begin{equation}
c_{mag}=A\tau^{B}(\cos\theta)^{C}
\end{equation}

\begin{figure*}
\includegraphics[width=1\linewidth]{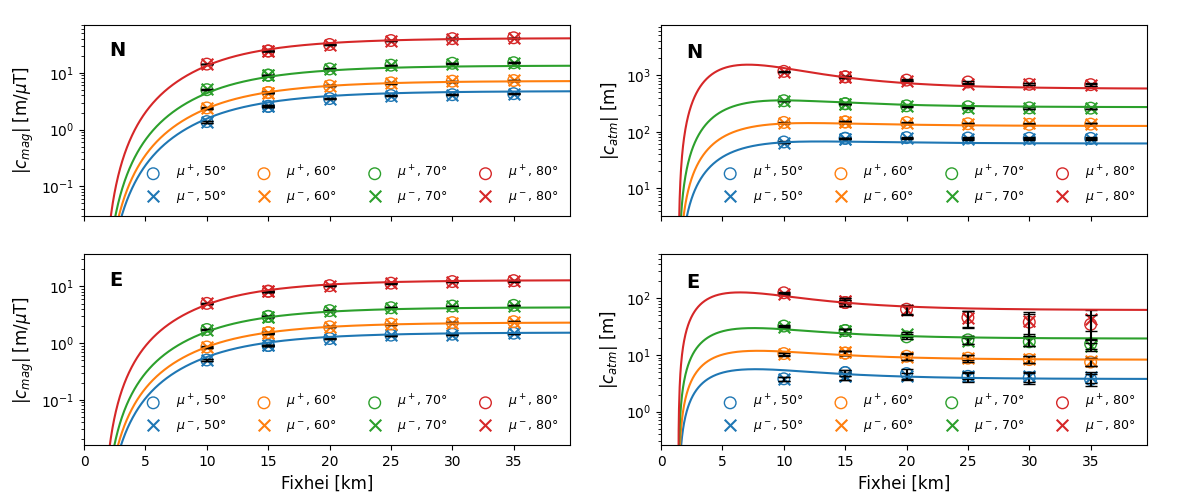}
\caption{Deflection parameters V.s fitting function for altitude of different first interaction points (FIXHEI) and zenith angles. The simulations are made with the parameters of the AUGER site. The deflection parameters for the overall deflection of particle number (N, upper) and energy (E, bottom) are presented respectively. }
\label{cmag_catm_merge}
\end{figure*}

Since the definition of $\tau$ contains $\cos(\theta)^{-1}$, when $B > C$, and $B>0$, the boundary conditions are satisfied.   Taking the site of the LHAASO site and the proton air showers as an example, as illustrated in the bottom panels of  Fig. \ref{cmag_catm_merge}, this fitting function matches the simulated results well — confirming robustness of the model.

%Next, we use this expression to fit the simulation results. In Section \ref{Validation of hypothesis}, we proved that $c_{mag}(\mu^+)\simeq -c_{mag}(\mu^-)$, so we will use one set of parameters to simultaneously fit $|c_{mag}(\mu^+)|$ and $|c_{mag}(\mu^-)|$. The fitting parameters will vary depending on the observation location or the primary particle type. The values of the fitting parameters can be found in the Appendix. FIG.\ref{AUGER_cmag_p} takes AUGER with protons as the primary particles as an example.

%\begin{figure}[h]
%    \centering
%    \begin{subfigure}
%        \includegraphics[width=0.4\textwidth]{AUGER/cmag_N_p_fit.png}
%        \caption{$c_{mag}(N)$}
%        \label{}
%    \end{subfigure}
%    \begin{subfigure}
%        \includegraphics[width=0.4\textwidth]{AUGER/cmag_E_p_fit.png}
%        \caption{$c_{mag}(E)$}
%        \label{}
%    \end{subfigure}
%    \caption{Fitting curve for $c_{mag}$. Observation location is AUGER, and primary particle is proton with energy of 1e7 GeV.}
%    \label{AUGER_cmag_p}. 
%\end{figure}

%\emph{The boundary conditions and equations for $c_{atm}$}
Similarly, the boundary conditions of $c_{atm}$ can be expressed as:

\begin{equation}
\left\{\begin{matrix}
\theta \to 0^\circ, & |\Delta\vec{r}_{atm}|\to 0 & \mathrm{(i)}\\
\theta \to 90^\circ, & |\Delta\vec{r}_{atm}|\to \infty & \mathrm{(ii)} \\
\mathrm{FIXHEI}\to \mathrm{OBSLEV}, & |\Delta\vec{r}_{atm}|\to 0 & \mathrm{(iii)}\\
\mathrm{FIXHEI}\to 112km , & |\Delta\vec{r}_{atm}|> 0 & \mathrm{(iv)}
\end{matrix}\right.
\end{equation}

where condition (i) corresponds to the scenario in which a primary particle enters the atmosphere vertically, in which case the asymmetries induced by the related effects are suppressed.
%since there is no additional absorption or decay at different places on the ground plane (FIG.\ref{atmosphere_absorption}), $\Delta\vec r_{atm}=0$.
%We provide an expression that fits the simulations well
A suitable expression of $c_{atm}$ is given by:
\begin{equation}
c_{atm}=A \tau^B \tan\theta\ \exp(\frac{C\tau}{\tan\theta})
\end{equation}

When $B>0,\ C<0$, the boundary conditions are satisfied.  It is found that $c_{atm}(\mu^+)\simeq c_{atm}(\mu^-)$, as illustrated in the upper panels of FIG. \ref{cmag_catm_merge}.  It is found that this fitting function also matches the simulated results well, except for the overall energy deflection $\Delta\vec{r}_E$  of highly inclined air showers initiated at very high altitudes. In such cases, muon deflection is more strongly influenced by muon decay; furthermore, $\Delta\vec{r}_E$  is larger than in other scenarios because high-energy muons undergo greater deflection. However, the overall deflection in this case is dominated by the magnetic effect, so this discrepancy can be ignored.

%Since $c_{atm}$ is affected by both the atmosphere absorption and particle decay, the expression of $c_{atm}$ is complicated, and is harder to find the physical interpretation.

%In Section ref {Validation of hypothesis}, we also proved $c_{atm}(\mu^+)\simeq c_{atm}(\mu^-)$, so we will fit $c_{atm}(\mu^+)$ and $c_{atm}(\mu^-)$ simultaneously. 
The fitting parameters for different observation locations and primary particle types can be found in  Table \ref{tab:tableforall}.

\begin{figure}[htbp]
    \centering
%    \begin{subfigure}
        \includegraphics[width=0.5\textwidth]{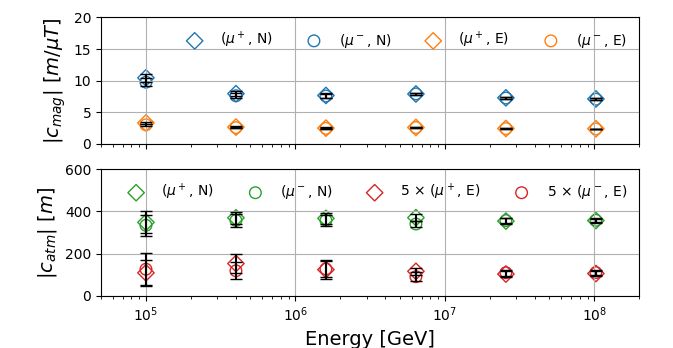}
%        \caption{$c_{mag}(N)$ and $c_{atm}(N)$}
%    \end{subfigure}
%    \begin{subfigure}
%        \includegraphics[width=0.4\textwidth]{otherfigure/c_energy_E.png}
%        \caption{$c_{mag}(E)$ and $c_{atm}(E)$}
%    \end{subfigure}
    \caption{$c_{mag}$ and $c_{atm}$ as function of primary  energy obtained in the simulations for the LHAASO site. The species, zenith, FIXHEI of primary particle is proton, 70$^\circ$, 20 km.}
    \label{LHAASO_energy}.
\end{figure}

%In Section \ref fitting c, we have validated the hypotheses and fitted the parameters. However, the energy of the primary particle is 1e7 GeV in all simulations. 
%In order to test the universality of the model, we will preliminarily verify that $c_{mag/atm}$ is approximately independent of the primary particle energy in the range of 1e5 GeV to 1e8 GeV.
%\emph{The relationship between ($c_{mag}$, $c_{atm}$) and primary energy. — } 

In order to test the dependence of ($c_{mag}$, $c_{atm}$) on primary energy, the analysis has been extended to the energy range from $10^{14}$ eV to $1.024  \cdot 10^{17}$ eV, with a step factor of four. As illustrated in FIG. \ref{LHAASO_energy}, the test was conducted with the parameters of the LHAASO site; the fitted deflection parameters ($c_{mag}$, $c_{atm}$) remained almost unchanged within this range, indicating the negligible dependence on primary energy.

\emph{Model uncertainties —} To evaluate the uncertainties of the deflection model, a relative error (used to measure the model's accuracy) is introduced as $ \varepsilon = \frac{|\Delta \vec{r}_{model}-\Delta \vec{r}_{simulation}|}{|\Delta \vec{r}_{simulation}|} $.  Using the parameters of the Pierre Auger observatory site, by comparing theoretical and simulated deflections, the distribution of the relative error for $\Delta\vec{r}_N$ $\Delta\vec{r}_E$ are presented by Fig. \ref{LHAASO_energy}, with 68\% confidence intervals for these errors are  11\% and 8\%, respectively — confirming the robustness of this model. This demonstrates that the bi-effect model can accurately describe the impact of the two effects on the behaviors of muons in air showers.
%When we predict the deflection of muons, there are two sources of error: the deflection curve fitting and the expression parameters A, B, C. We use relative error to measure the accuracy of the model fitting $ \varepsilon = \frac{|\Delta \vec{r}_{model}-\Delta \vec{r}_{simulation}|}{|\Delta \vec{r}_{simulation}|} $

\begin{figure}[htbp]
    \centering
        \includegraphics[width=0.4\textwidth]{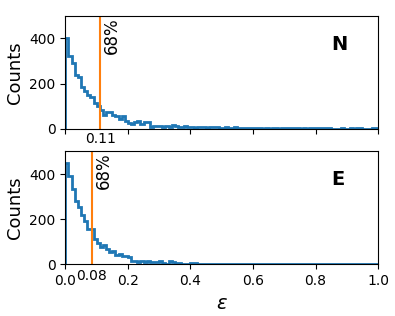}
        %\caption{$\Delta\vec{r}_N$}
    \caption{Relative error distribution of the deflection model.   1728 air showers (6912 deflection data) with energy of $10^7$ GeV are simulated for site of the Pierre Auger observatory.}
    \label{relativeerror}.
\end{figure}

\emph{Simple applications —}
%\emph{The meaning of model parameters-} 
Despite the complex physical processes involved in air shower development, the bi-effect model provides a very simple and clear explanation of how atmospheric and magnetic effect influence the muons' movement.  The simplest application of the bi-effect model is its straightforward use to predict the overall deflections of particle number and energy across different experiments. The detection capability of dense arrays — such as LHAASO and SKA — is limited by their relatively small effective area, which prevents them from reconstructing air showers with large zenith angles. The challenges posed by such deflections can be addressed by adding detector units at the site. The dual-effect model facilitates detector upgrades at the lowest cost: by placing least new detectors around the core of the muon distribution.

\begin{figure}[htbp]
    \centering
    \includegraphics[width=0.3\textwidth]{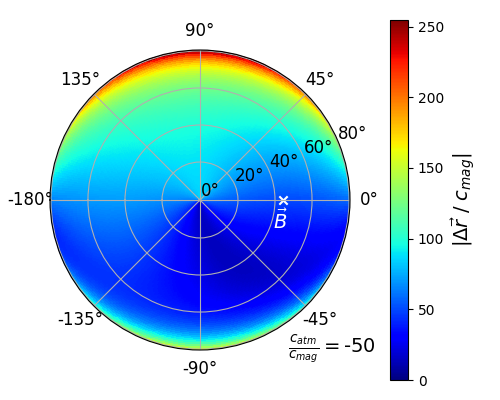}
    \caption{The magnitude relationship between $|\Delta{\vec{r}_{mag}}|$ and $c_{mag}$ (while $\frac{c_{atm}}{c_{mag}}=-50$). The radial and angular directions correspond to the zenith angle and azimuth angle, respectively. The unit of $\Delta{\vec{r}}$ and $c_{mag}$ is $[m]$, $[m/\mu T]$.(LHAASO site)}
    \label{r_cmag}
\end{figure}

An example is shown in FIG. \ref{r_cmag}, for the LHAASO site, to detect more inclined cosmic rays, detectors should be deployed over a larger area to the west of the original array, whereas the area requiring deployment to the east can be smaller. If the value of the deflection parameters are known,  the deflection for different shower directions can be derived directly by multiplying the magnitude by $c_{mag}$.  Based on this result, when an experiment requires an upgrade to detect cosmic rays within a specific large zenith angle range, the precise area for deploying new detector units can be directly derived from this result.

As shown in Table \ref{tab:tableforall}, for an experimental site, the fitting parameter values differ for different primary particles. These parameters can therefore be used to determine the mass composition of cosmic rays. This approach will be investigated in details in a separate paper.

%\begin{figure}[htbp]
%    \centering
%    \includegraphics[width=0.3\textwidth]%{r_cmag/r_cmag_800_LHAASO_polar.png}
%    \caption{The magnitude relationship between $|\Delta{\vec{r}}|$ and $c_{mag}$ when considering $\Delta{\vec{r_{atm}}}$. When $c_{mag}|\vec{B}|\sim c_{atm}$, the symmetry is broken, and when $c_{mag}|\vec{B}|\ll  c_{atm}$, atmosphere absorption dominates deflection. ( LHAASO, $\frac{c_{atm}}{c_{mag}}=-800$) }
%    \label{r_cmag_catm}
%\end{figure}

\section{Conclusions and discussions}
%\emph{Conclusion and discussions —}
In this work, we attempt to describe the deflection phenomenon of secondary particles in extensive air showers. Based on the two fundamental hypotheses, we derived the expression for the deflection model through mathematical derivation, which has a clear and concise physical interpretation.

Subsequently, through air shower simulations, we validated the hypotheses and confirmed that the model can accurately fit the deflection of secondary particles.  This attempt has successfully integrated the deflections induced by atmospheric effects and the Lorentz force. 

Owing to the muon puzzle and the energy threshold constrained by the local magnetic field, this model needs to be calibrated with real data to obtain more accurate fitting parameters. Additionally, other factors may exist that could further improve the model.

Next, based on the boundary conditions defined by the model, we obtained two empirical formulas for $c_{mag}$ and $c_{atm}$ as functions of the primary particle physical properties through  cosmic ray simulations. This may enable the reconstruction of primary particle information, including mass composition. This analysis was conducted for three sites, demonstrating that the method is generic and can be applied to any relevant experiment.

This model does not exhibit an obvious dependence on primary energy; however, the simulations in this work do not cover ultra-high energies, so this aspect may require further verification.

For electromagnetic particles, their high energy loss rate in the atmosphere introduces excessive uncertainties into the analysis, making it difficult to model this effect. The corresponding results will be presented in a separate paper.

\section{Acknowledgments}
We thank Ying-Ying Guo, Yi Zhang, Shi-ping Zhao, Ke-wen Zhang for the discussions. This work is supported by the National Science Foundation of China under grants No.12393852 and No.12333006.

\appendix
\section{Deflection curve} 
\begin{figure}[htbp]
    \centering
    \includegraphics[width=0.7\linewidth]{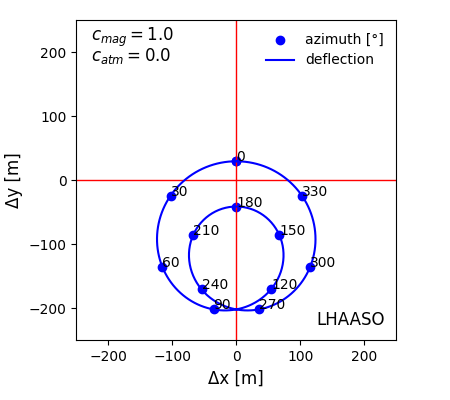}
    \caption{The deflection curve in different azimuth angle (in ground plane). $\theta$ is the zenith angle of primary particle. Taking the geomagnetic field of LHAASO as an example. (LHAASO, $\theta=80^\circ$)}
    \label{DC}
\end{figure}

Since we have assumed that $c_{mag}$ and $c_{atm}$ are independent of the primary particle's incident azimuth, we can plot the deflection curves for different azimuth angles using the same set of ($c_{mag}$, $c_{atm}$).

We begin by plotting the case where $c_{atm}=0$ (as shown in FIG. \ref{DC}). $\Delta\vec{r}$ increases rapidly as the zenith angle grows. When the incident zenith angle of the particle exceeds the zenith angle of the magnetic field, the deflection curve exhibits a double-circle structure. Furthermore, the sizes of the two circles gradually converge as the zenith angle increases.

%\section{Fitting parameters}  
\begin{table}
\caption{Parameter table for $c_{mag}$ and $c_{atm}$. For $\mu^+$, $A_{mag}>0$, and for $\mu^-$, $A_{mag}<0$.}
\label{tab:tableforall}
\begin{ruledtabular}
\begin{tabular}{cccc}
%& $ \ \  $ &\multicolumn{1}{c}{$c_{mag}$}&\multicolumn{4}{c}{$c_{atm}$}\\
$\Delta \vec r$ [m] &$|A_{mag}|$& $B_{mag}$ &$C_{mag}$ ($[m/\mu T]$) \\ \hline
$\Delta \vec r_N$(AUGER, p)&$3.463\pm0.010$&$3.024\pm0.007$&$1.381\pm0.007$\\
$\Delta \vec r_E$(AUGER, p)&$1.075\pm0.003$&$2.682\pm0.007$&$1.056\pm0.007$\\
$\Delta \vec r_N$(AUGER, Fe)&$3.991\pm0.009$&$3.046\pm0.006$&$1.470\pm0.006$\\
$\Delta \vec r_E$(AUGER, Fe)&$1.197\pm0.002$&$2.701\pm0.006$&$1.122\pm0.006$\\
$\Delta \vec r_N$(LHAASO, p)&$5.802\pm0.026$&$2.860\pm0.006$&$0.994\pm0.006$\\
$\Delta \vec r_E$(LHAASO, p)&$1.810\pm0.006$&$2.605\pm0.005$&$0.803\pm0.005$\\
$\Delta \vec r_N$(LHAASO, Fe)&$6.905\pm0.025$&$2.891\pm0.005$&$1.080\pm0.005$\\
$\Delta \vec r_E$(LHAASO, Fe)&$1.995\pm0.006$&$2.568\pm0.004$&$0.805\pm0.004$\\
$\Delta \vec r_N$(SKA, p)&$2.795\pm0.005$&$3.026\pm0.007$&$1.470\pm0.007$\\
$\Delta \vec r_E$(SKA, p)&$0.867\pm0.001$&$2.703\pm0.007$&$1.127\pm0.007$\\
$\Delta \vec r_N$(SKA, Fe)&$3.122\pm0.004$&$3.099\pm0.005$&$1.584\pm0.005$\\
$\Delta \vec r_E$(SKA, Fe)&$0.949\pm0.001$&$2.716\pm0.005$&$1.174\pm0.005$\\
\hline

$\Delta \vec r$[m] & $A_{atm}$ & $B_{atm}$ &$C_{atm} ([m])$\\ \hline

$\Delta \vec r_N$(AUGER, p)&$-130.4\pm1.0$&$1.768\pm0.005$&$-1.013\pm0.007$\\
$\Delta \vec r_E$(AUGER, p)&$-18.3\pm0.6$&$1.362\pm0.013$&$-1.705\pm0.036$\\
$\Delta \vec r_N$(AUGER, Fe)&$-137.4\pm0.9$&$1.751\pm0.004$&$-1.060\pm0.006$\\
$\Delta \vec r_E$(AUGER, Fe)&$-19.8\pm0.5$&$1.385\pm0.010$&$-1.761\pm0.031$\\
$\Delta \vec r_N$(LHAASO, p)&$-158.5\pm0.6$&$2.056\pm0.006$&$-1.232\pm0.005$\\
$\Delta \vec r_E$(LHAASO, p)&$-6.3\pm0.1$&$2.797\pm0.024$&$-0.654\pm0.012$\\
$\Delta \vec r_N$(LHAASO, Fe)&$-176.0\pm0.6$&$1.999\pm0.005$&$-1.358\pm0.005$\\
$\Delta \vec r_E$(LHAASO, Fe)&$-7.3\pm0.1$&$2.706\pm0.019$&$-0.732\pm0.012$\\
$\Delta \vec r_N$(SKA, p)&$-109.3\pm1.7$&$1.753\pm0.008$&$-0.855\pm0.011$\\
$\Delta \vec r_E$(SKA, p)&$-14.8\pm0.8$&$1.492\pm0.022$&$-1.259\pm0.051$\\
$\Delta \vec r_N$(SKA, Fe)&$-114.0\pm1.6$&$1.756\pm0.007$&$-0.878\pm0.010$\\
$\Delta \vec r_E$(SKA, Fe)&$-19.1\pm0.9$&$1.465\pm0.015$&$-1.465\pm0.043$\\
\end{tabular}
\end{ruledtabular}
\end{table}

% The \nocite command causes all entries in a bibliography to be printed out
% whether or not they are actually referenced in the text. This is appropriate
% for the sample file to show the different styles of references, but authors
% most likely will not want to use it.
\nocite{*}
\newpage
\bibliography{main}% Produces the bibliography via BibTeX.

\end{document}